\documentclass[12pt]{article}

\def\fnote#1#2{\begingroup\def\thefootnote{#1}\footnote{#2}\addtocounter{footnote}{-1}\endgroup}

\usepackage{amssymb}

\def\inbar{\vrule height1.5ex width.4pt depth0pt}
\def\IB{\relax{\rm I\kern-.18em B}}
\def\IC{\relax\,\hbox{$\inbar\kern-.3em{\rm C}$}}
\def\ID{\relax{\rm I\kern-.18em D}}
\def\IE{\relax{\rm I\kern-.18em E}}
\def\IF{\relax{\rm I\kern-.18em F}}
\def\IG{\relax\,\hbox{$\inbar\kern-.3em{\rm G}$}}
\def\IH{\relax{\rm I\kern-.18em H}}
\def\II{\relax{\rm I\kern-.18em I}}
\def\IK{\relax{\rm I\kern-.18em K}}
\def\IL{\relax{\rm I\kern-.18em L}}
\def\IM{\relax{\rm I\kern-.18em M}}
\def\IN{\relax{\rm I\kern-.18em N}}
\def\IO{\relax\,\hbox{$\inbar\kern-.3em{\rm O}$}}
\def\IP{\relax{\rm I\kern-.18em P}}
\def\IQ{\relax\,\hbox{$\inbar\kern-.3em{\rm Q}$}}
\def\IR{\relax{\rm I\kern-.18em R}}
\def\IT{\relax{\rm I\kern-.18em T}}
\def\ZZ{\relax{\sf Z\kern-.4em Z}}

\def\a{\alpha}   \def\b{\beta}    \def\g{\gamma}  
\def\e{\epsilon}      
    \def\Om{\Omega} \def\si{\sigma}

\def\cA{{\cal A}} 
   
 \def\cH{{\cal H}} \def\cI{{\cal I}} 
   
\def\cO{{\cal O}} \def\cP{{\cal P}}

\def\afrak{{\mathfrak a}} 
 \def\pfrak{{\mathfrak p}}

\def\mathC{{\mathbb C}}  \def\mathF{{\mathbb F}}
  \def\mathP{{\mathbb P}} \def\mathQ{{\mathbb Q}}
  \def\mathZ{{\mathbb Z}}

      \def\bX{{\bar X}}

\def\fnote#1#2{\begingroup\def\thefootnote{#1}\footnote{#2}\addtocounter
{footnote}{-1}\endgroup}
\def\beq{\begin{equation}}
\def\eeq{\end{equation}}
\def\bea{\begin{eqnarray}}
\def\eea{\end{eqnarray}}
\def\llea#1{\label{#1}\eea}
\def\lleq#1{\label{#1}\eeq}
\let\nn=\nonumber
\def\tabroom{\hbox to0pt{\phantom{\Huge A}\hss}}
\def\notin{\ \hbox{{$\in$}\kern-.51em\hbox{/}}}

\def\lra{\longrightarrow}

\def\ra{{\rightarrow}}

  \def\E1Fq{E_1/\IF_q}

\def\oK{{\overline K}}

\def\omathQ{{\overline \mathQ}}

\def\un{{\underline n}}

\def\ormF{{\overline{\rmF}}}

\def\rmA{{\rm A}}  \def\rmB{{\rm B}}    \def\rmF{{\rm F}}
\def\rmH{{\rm H}}  \def\rmK{{\rm K}}    \def\rmN{{\rm N}}
\def\rmT{{\rm T}}

     \def\rmdeg{{\rm deg}}
\def\rmdet{{\rm det}}   \def\rmdim{{\rm dim}}
        
   \def\rmmod{{\rm mod}}   \def\rmrk{{\rm rk}}
 
\def\rmsign{{\rm sign}} \def\rmth{{\rm th}}     
\def\rmtr{{\rm tr}}

\def\rmAut{{\rm Aut}}       
   \def\rmFr{{\rm Fr}}    \def\rmGal{{\rm Gal}}
\def\rmGL{{\rm GL}}       
\def\rmHW{{\rm HW}}       \def\rmNS{{\rm NS}}
     \def\rmRe{{\rm Re}}
\def\rmSL{{\rm SL}}     \def\rmSpec{{\rm Spec}} 
\def\rmTr{{\rm Tr}}

\def\rmK3{{\rm K3}}

\def\notdiv{{\relax{~|\kern-.35em /~}}}

\def\boxit#1{
\vbox{\hrule height1pt\hbox{\vrule width1pt\kern0.3cm
\vbox{\kern0.3cm\hbox{$\displaystyle#1$}\kern0.3cm}\kern0.3cm\vrule
width1pt}\hrule height1pt}}

\begin{document}
\parindent=0pt
\hfill {\bf NSF$-$KITP$-$06$-$23}


\vskip 0.8truein

 \centerline{\large {\bf The Langlands Program and String Modular K3 Surfaces}}


\vskip .1truein

\vskip 0.3truein

\centerline{{\sc Rolf Schimmrigk}\fnote{$\dagger$}{email:
netahu@yahoo.com}}

\vskip .3truein

\centerline{\it Indiana University South Bend}
 \vskip .05truein
\centerline{\it 1700 Mishawaka Ave., South Bend, IN 46634}

\vskip 1truein

\baselineskip=19pt

\centerline{\bf ABSTRACT:}

\vskip .2truein

\begin{quote}
 A number theoretic approach to string compactification is developed
 for Calabi-Yau hypersurfaces in arbitrary dimensions. The motivic
 strategy involved is illustrated by showing that the Hecke eigenforms
 derived from Galois group orbits of the holomorphic two-form of a
 particular type of K3 surfaces can be expressed in terms of modular
 forms constructed from the worldsheet theory. The process of deriving
 string physics from spacetime geometry can be reversed, allowing the
 construction of K3 surface geometry from the string characters of the
 partition function. A general argument for K3 modularity follows
 from mirror symmetry, in combination with the proof of the
 Shimura-Taniyama conjecture.
\end{quote}

\vfill

{\sc PACS Numbers and Keywords:} \hfill \break Math:  11G25
Varieties over finite fields; 11G40 L-functions; 14G10  Zeta
functions; 14G40 Arithmetic Varieties \hfill \break Phys: 11.25.-w
Fundamental strings; 11.25.Hf Conformal Field Theory; 11.25.Mj
Compactification

\renewcommand\thepage{}
\newpage

\pagenumbering{arabic}

 \tableofcontents

\vfill \eject



\baselineskip=17.2pt
\parskip=.2truein
\parindent=0pt

\section{Introduction}

One of the intriguing aspects of string theory is the possibility
of understanding the structure of spacetime from first principles
in terms of the physics of the worldsheet. In the past a number of
different techniques have been used to address this question, e.g.
Landau-Ginzburg theories and non-linear sigma models. The aim of
the present paper is to continue a different program that uses
methods from arithmetic geometry to understand this problem in the
context of exactly solvable Calabi-Yau varieties. The structure of
higher dimensional Calabi-Yau varieties is quite intricate and
there are several aspects of these theories that lend themselves
to an arithmetic analysis. The focus here will be on formulating a
framework that combines methods from algebraic number theory and
arithmetic geometry in the context of Calabi-Yau hypersurfaces of
arbitrary dimensions. This general approach is then applied to a
special class of K3 surfaces by showing that basic building blocks
of the underlying string partition functions can be derived from
the geometry of these K3s.

The simplest case in which the idea of using arithmetic geometry
to derive worldsheet information from geometry can be tested is
provided by the framework of toroidal compactifications. Elliptic
curves provide useful examples because of the proof of the
Shimura-Taniyama conjecture. This modularity theorem states that
all elliptic curves defined over the rational number field are
modular in the sense that the Mellin transform of their Hasse-Weil
L-series defines a modular form of weight two with respect to some
congruence subgroup of $\rmSL(2,\mathZ)$ \cite{w95, d96, cdt99,
bcdt01}. In the context of exactly solvable elliptic curves the
issue then becomes whether the modular forms derived from these
curves can be expressed in terms of modular forms derived from the
underlying superconformal field theory. It was shown in refs.
 \cite{su02,ls04,s05} that this is possible in terms of cusp
 forms constructed from the string functions associated to the
 affine Lie algebra
 $A^{(1)}_1$ of the $N=2$ superconformal minimal models.

The generalization of this result to higher genus curves and
higher dimensional varieties is made difficult by the fact that no
analog of the elliptic modularity theorem is known, even
conjecturally. There is a general expectation, often summarized as
part of the Langlands program \cite{l78}, that many varieties will
lead to modular forms, but there is no known systematic procedure
which provides guidance how this should be accomplished. There is
in particular no known higher dimensional analog of Weil's
experimental observation \cite{w67} concerning the geometric
interpretation of the level of a modular form in the context of
elliptic curves, an ingredient that galvanized interest in the
Shimura-Taniyama conjecture, culminating in Wiles' breakthrough.
For higher genus curves of Brieskorn-Pham type the situation is a
little better because one can use a known factorization procedure
to decompose the associated Jacobian variety into simple abelian
factors, which then can be tested for modularity \cite{ls04}.

In general the L-function of a variety will not be an interesting
object in the context of recovering string theoretic modular
forms. Except for particularly simple varieties, such as rigid
Calabi-Yau threefolds, for which many modular forms have been
identified (ref. \cite{y03} contains articles with many further
references), the L-function itself does not lead to a modular
form. It is known, from Grothendieck's proof \cite{g65} of part of
the Weil conjectures \cite{w49}, that Artin's congruent zeta
function factorizes in a way that is determined by the cohomology
groups of the variety. The zeta function at a prime $p$ decomposes
into the quotient
 \beq
    Z(X/\mathF_p,t) =
    \frac{\cP_p^{1}(t)\cP^{3}_p(t)\cdots \cP^{2n-1}_p(t)}{
   \cP_p^{0}(t)\cP^{2}_p(t)\cdots \cP^{2n}_p(t)},
   \lleq{ratz}
 where $\rmdim_{\mathC}X=n$, and $\cP^{i}_p(t)$ is a polynomial
  \beq
  \cP^{i}_p(t) = \sum_{j=0}^{b^i} \b^i_j(p)t^j
  \eeq
 associated to the $i^{\rmth}$ cohomology group,
  with a degree $b^i = \rmdim~ \rmH^i(X)$ given by the
 $i^{\rmth}$ Betti number. This result motivates the
 introduction of L$-$functions associated to the individual cohomology
 groups, thereby reducing the complexity of the zeta function.
 Even though this factorization provides an important simplification, it is not
 enough for string theory. The individual cohomology groups can be
 quite complicated because the Betti numbers of Calabi-Yau
 varieties tend to be large. The idea in this paper is to factorize
 these polynomials further,  and to consider  L$-$functions associated to
 the resulting factors. The problem that arises is that it is unclear
 a priori which type of factorization leads to a physically
 meaningful L-function.

In the absence of a clear understanding of what the conditions are
in higher dimensions that can lead to string theoretic modular
forms on the worldsheet, it is useful to identify  selection rules
that guide the factorization of the L-functions. There are several
ways to think about this problem, and in the present paper the
following point of view will be adopted. The first idea is
utilitarian in nature, guided by the expectation that the results
of \cite{su02, ls04, s05}, or some not too radical modification
thereof, will generalize to higher dimensions. It follows from
those results that the string theoretic modular forms relevant for
the arithmetic approach are forms with coefficients that are
rational integers. The primitive factors of the polynomials
$\cP_p^i(t)$ that arise from the complete factorization
 \beq
   \cP_p^{i}(t) = \prod_{j=1}^{b^i}(1-\g^i_j(p)t),
  \eeq
  lead to L-function factors with coefficients $\g^i_j(p)$ that are
  algebraic integers in number fields, defined by extensions of the
  rational number field $\mathQ$.  This shows that a complete
factorization is not useful in the present context, and that from
a practical point of view one should be guided by the idea of
identifying pieces of the cohomology that lead to forms with
coefficients in $\mathZ$.

A more conceptual way of thinking about this problem is
representation theoretic. Associated to a Calabi-Yau hypersurface
in a toric variety is a cyclotomic number field, which for
Brieskorn-Pham spaces admits an interpretation as the fusion field
of the underlying conformal field theory \cite{s03}. The Galois
group of this field is a finite cyclic group which acts on the
cohomology of the variety. The action of this group is reducible
in general, and therefore one can use it to decompose the
cohomology group into pieces defined by the irreducible
representations of the group. A possible strategy therefore is to
focus on the L-functions associated to these irreducible
representations of the Galois group. A distinguished element in
the cohomology ring of any $n-$dimensional Calabi-Yau variety is
the holomorphic $n-$form, leading to the concept of an
$\Om-$motive of a Calabi-Yau variety. The notion of such a motive
is general, and the question addressed here is whether this motive
is string modular in some sense. The approach formulated therefore
provides a general method to gain a better understanding of the
relation between the geometry of spacetime in string theory and
the physics on the worldsheet.

In the present paper the strategy just described is illustrated by
considering the class of extremal K3 surfaces of Brieskorn-Pham
type, i.e. surfaces $S$ which over the complex field $\mathC$ have
maximal Picard number $\rho(S)=20$. These surfaces are of the form
 \bea
   S^4 &=& \left\{(z_0:\cdots :z_3)\in \mathP_3~{\Big
   |}~ z_0^4 +z_1^4+z_2^4+z_3^4 = 0 \right\}, \nn \\
   S^{6\rmA} &=& \left\{(z_0:\cdots :z_3)\in \mathP_{(1,1,1,3)}~{\Big
   |}~ z_0^6+z_1^6 + z_2^6+z_3^2=0 \right\}, \nn \\
   S^{6\rmB} &=& \left\{(z_0:\cdots :z_3)\in \mathP_{(1,1,2,2)}~{\Big
   |}~ z_0^6+z_1^6 + z_2^3+z_3^3=0 \right\}.
 \llea{examples}

In order to state the results of this analysis some notation is
needed. With $q=e^{2\pi i \tau}$, let $f(q)=\sum_n a_nq^n$ be a
cusp form, and
 $\vee^2 f(q) = \sum_n b_n q^n$ be the product given by $b_p = a_p^2
-2p$. The motivation for this definition will become clear below.
Define the Hecke congruence subgroup $\Gamma_0(N)$ as
 \beq
 \Gamma_0(N) = \left\{\left(\matrix{a&b\cr
c&d\cr}\right) \in \rmSL(2,\mathZ) ~{\Big |}~\left(\matrix{a&b\cr
c&d\cr}\right) \sim \left(\matrix{*&*\cr 0&*\cr}\right)
~(\rmmod~N)\right\},
 \eeq
 and denote the Galois group of the cyclotomic number field
 $\mathQ(\mu_d)$ by $\rmGal(\mathQ(\mu_d)/\mathQ)$, were $\mu_d$
 is the cyclic group generated by
 a primitive $d^{\rmth}$ root of unity. The Dedekind eta
 function is given by $\eta(q)=q^{1/24}\prod_{n=1}^{\infty} (1-q^n)$,
 while the theta functions
   $\Theta^k_{\ell,m}(\tau) = \eta^3(\tau)c^k_{\ell,m}(\tau)$
 are Hecke indefinite modular forms
 associated to the Kac-Peterson string functions $c^k_{\ell,m}(\tau)$ of the
 affine algebra $A_1^{(1)}$ at level $k$. Finally, the quadratic
 characters determined by the Legendre symbol are written as
 $\chi_n(p) = (\frac{n}{p})$. $\vartheta(q)$ is a modular
 form of weight one described further below. The following results
 will be shown.

{\bf Theorem 1.}
 {\it Let $M_{\Om} \subset H^2(S^d)$ be the irreducible representation
 of $\rmGal(\mathQ(\mu_d)/\mathQ)$ associated to the holomorphic 2$-$form
 $\Om \in H^{2,0}(S^d)$ of the K3 surface $S^d$, where $d=4,6\rmA, 6\rmB$.
 Then the Mellin transforms $f_{\Om}(S^d,q)$
 of the L-functions $L_{\Om}(S^d,s)$ associated to $M_{\Om}$ are
 given by}
  \bea
   f_{\Om}(S^4,q) &=& \eta^6(q^4) \nn \\
   f_{\Om}(S^{6\rmA},q) &=& \vartheta(q^3) \eta^2(q^3)\eta^2(q^9) \nn \\
   f_{\Om}(S^{6\rmB},q) &=& \eta^3(q^2)\eta^3(q^6) \otimes \chi_3.
  \llea{k3mod}
 {\it These functions are cusp forms of weight three with respect
 to $\Gamma_0(N)$ with levels $16, 27$ and 48, respectively.
 For $S^4$ and $S^{6\rmA}$ the L-functions can be written as
  $L_{\Om}(S^d,s) = L(\vee^2 f_d, s)$, where $f_d(q)$ are cusp forms
  of weight two and levels 64 and 27, respectively, given by}
   \bea
    f_4(\tau) &=& \Theta^2_{1,1}(4\tau)^2\otimes \chi_2 \nn \\
    f_{6\rmA}(\tau) &=&
    \Theta^1_{1,1}(3\tau)\Theta^1_{1,1}(9\tau).
   \eea
  {\it For $S^{6\rmB}$ the L-series is given by
   $L_{\Om}(S^{6\rmB},s) = L(\vee^2 f_{6\rmB}\otimes \chi_3,s)$ in terms of
   the cusp form of level 144}
   \beq
    f_{6\rmB}(\tau) = \Theta^1_{1,1}(6\tau)^2\otimes \chi_3.
   \eeq

Physically, this result proves a string theoretic interpretation
of the motivic L-function associated to the holomorphic $\Om-$form
of K3 surfaces in terms of the affine Lie algebra $A_1^{(1)}$ on
the worldsheet. It generalizes results in lower dimensions for the
Hasse-Weil L-function of elliptic Brieskorn-Pham curves obtained
in \cite{su02,ls04,s05}. Mathematically, it can be viewed as
providing a motivic interpretation of modular forms derived from
Kac-Moody algebras, i.e. it provides a string theoretic origin of
modular motives for a class of K3s.

{\bf Corollary.} {\it The $\Om-$motivic modular forms of
 extremal K3 surfaces of Brieskorn-Pham type are twisted products of Hecke
indefinite modular forms that arise from Kac-Peterson string
functions.}

The modular forms of weight two that appear in  the theorem as
building blocks of the arithmetic structure of extremal
Brieskorn-Pham K3 surfaces are all elliptic, i.e. the Mellin
transforms of Hasse-Weil L-series of elliptic curves. Consider the
class of elliptic Brieskorn-Pham curves given by
  \bea
 E^3 &=& \left\{(z_0:z_1:z_2) \in \IP_2~|~z_0^3+z_1^3+z_2^3=0
            \right\} \nn \\
 E^4 &=& \left\{(z_0:z_1:z_2) \in \IP_{(1,1,2)}~|~z_0^4+z_1^4+z_2^2=0
          \right\} \nn \\
 E^6 &=& \left\{(z_0:z_1:z_2) \in \IP_{(1,2,3)}~|~z_0^6+z_1^3+z_2^2=0 \right\},
 \eea
 All three curves are string modular in the following sense \cite{s05}.

{\bf Theorem 2.}~{\it The Mellin transforms $f_{\rmHW}(E^d,q)$ of
the Hasse-Weil L-functions $L_{\rmHW}(E^d,s)$ of the curves $E^d,
d=3,4,6$ are modular forms $f_{\rmHW}(E^d,q) \in
S_2(\Gamma_0(N))$, with $N\in \{27, 64, 144\}$, respectively.
These forms factor into products of Hecke indefinite modular forms
as follows}
 \bea
f_{\rmHW}(E^3,q) &=& \Theta^1_{1,1}(q^3)\Theta^1_{1,1}(q^9) \nn \\
f_{\rmHW}(E^4,q) &=& \Theta^2_{1,1}(q^4)^2\otimes \chi_2 \nn \\
f_{\rmHW}(E^6,q) &=& \Theta^1_{1,1}(q^6)^2\otimes \chi_3.
 \llea{hw-as-thetas}
 This shows that the string theoretic nature of the modular forms
 of extremal Brieskorn-Pham K3 surfaces is induced by the string
 theoretic modularity of elliptic Brieskorn-Pham curves.

Given the central nature of modularity in string theory, it is
natural to ask how general modularity is for K3 surfaces. An
argument that points to modularity as a common property can be
made by combining mirror symmetry with the elliptic modularity
theorem of \cite{bcdt01}. The original explicit observation of
mirror symmetry \cite{cls90, gp90} has been interpreted in a
number of different ways, e.g. in terms of toric geometry by
Batyrev \cite{b94}, in a homological context  by Kontsevich
\cite{k94}, and in terms of fibrations by Strominger, Yau and
Zaslow \cite{syz95}. It is the latter framework that is most
useful for K3 modularity. The idea of the SYZ conjecture is based
on a toroidal fibration structure of general Calabi-Yau varieties
that is suggested by D-branes. For complex dimension two this
conjecture implies that mirror pairs of K3 surfaces are
characterized by fibrations in terms of elliptic curves. For
elliptic curves defined over $\mathQ$ the modularity theorem
proven by Breuil, Conrad, Diamond and Taylor shows that every
elliptic curve is modular in the sense that the Mellin transform
of its associated Hasse-Weil L-function is a modular form of
weight 2 with respect to a congruence group that is determined by
the conductor of the curve. The SYZ conjecture and the modularity
theorem therefore imply that mirror pairs of K3 surfaces defined
over $\mathQ$ are modular.

The outline of the paper is as follows. Section 2 contains the
arguments for considering irreducible Galois representations as
the modular building blocks of Calabi-Yau varieties. These
arguments are general, not restricted to K3 surfaces. Section 3
computes the L-series of the extremal K3 surfaces of
Brieskorn-Pham type. Section 4 leads to the identification of the
modular forms derived from these K3 surfaces with modular forms
derived from the string worldsheet, completing the identifications
of the theorem. Section 5 complements the previous computations by
constructing the K3 surfaces via the twist map \cite{hs99}. The
philosophy adopted here is similar to the one used in \cite{ls04}
in the context of higher genus curves. Section 6 describes the
identification of a singular K3 surface in terms of string
theoretic modular forms. Section 7 generalizes to K3 surfaces
aspects of complex multiplication and arithmetic moonshine,
discussed in \cite{s05} in the context of elliptic curves. The
appendix proves the identification of the geometric and the
modular L-series to all orders. The proof is based on the method
of Faltings-Serre-Livn\'e, and uses results from the
representation theory of the absolute Galois group
$\rmGal(\omathQ/\mathQ)$ of the rational numbers.

\vskip .1truein

\section{$\Om-$Motivic L-functions of Calabi-Yau varieties}

\subsection{Counting and Jacobi sums}

L-functions of projective varieties can be computed in several
ways. The most direct procedure starts from the rational form
(\ref{ratz}) of Artin's congruent zeta function
  \beq
  Z(X/\mathF_p,t) =
 \exp\left(\sum_{r\geq 1} \#(X/\mathF_{p^r}) \frac{t^r}{r}
 \right),
 \eeq
 defined in terms of the degree $r$ extensions $\mathF_{p^r}$ of the
 finite field $\mathF_p$ for any prime $p$.
 This factorization of the zeta function leads to the definition
 of cohomological L-functions
 \beq
  L^{(i)}(X,s) =
  \prod_{p~{\rm prime}} \frac{1}{\cP^i_p(p^{-s})}
  \lleq{lfunction}
associated to the $i^{\rmth}$ cohomology group $H^i(X)$.

The simplest way of computing this function is via a direct count
of the number of solutions $N_{r,p}=\#(X/\mathF_{p^r})$. For the
case of K3 surfaces the cohomological form of the zeta function
simplifies to
 \beq
  Z(X/\mathF_q,t) = \frac{1}{(1-t)\cP^2_p(t)(1-p^2t)}.
 \eeq
 Expanding this quotient as
 \bea
 Z(X/\mathF_p,t)
&=& 1 + (p^2+1-\beta_1)t
  +\left[p^4 +p^2(1-\beta_1)+\beta_1^2-\beta_1 -\beta_2+1\right]t^2 +
  \cdots,
 \eea
 and comparing it to the expansion of the exponential, leads to
 \bea
 \beta_1^2(p) &=& 1+p^2-N_{1,p} \nn \\
 \beta_2^2(p) &=& 1+p^2+p^4 - 3(1+p^2)N_{1,p} +
  \frac{1}{2}\left(N_{1,p}^2-N_{2,p}\right) \nn \\
  &\vdots &
 \eea
 for the polynomial $\cP^2_p(t) = \sum_i \b^2_i(p)t^i$.

For weighted Fermat varieties a second method for obtaining both
the cardinalities and the L-function was introduced by Weil
\cite{w49}. This formulation is particularly useful because it
allows to disentangle the complicated cohomological structure of
higher dimensional varieties in a systematic way, at least for
this special type \cite{w52}.

{\bf Theorem 3.}  {\it For a smooth weighted projective surface
with degree vector $\un=(n_0,...,n_s)$}
 \beq
 X^{\un} =\{z_0^{n_0}+z_1^{n_1} + \cdots + z_s^{n_s} =0 \}
  \subset \mathP_{(k_0,k_1,\dots,k_s)}
   \eeq
  {\it defined over the finite field $\mathF_q$ define the set
  $\cA_s^{q,\un}$ of rational vectors $\a =(\a_0,\a_1,\dots,\a_s)$ as}
  \beq
  \cA_s^{q,\un} = \left\{\a \in \mathQ^{s+1} ~|~ 0<\a_i<1, ~d_i=(n_i,q-1),
  ~d_i\a_i =0~\rmmod~1,~\sum_{i=0}^s \a_i=0~(\rmmod~1) \right\}.
  \eeq
  {\it For each $(s+1)-$tuple $\a$ define the Jacobi sum }
 \beq
 j_q(\a_0,\a_1,\dots,\a_s) = \frac{1}{q-1}
  \sum_{\stackrel{u_i \in \mathF_q}{u_0+u_1+\cdots +u_s=0}}
   \chi_{\a_0}(u_0) \chi_{\a_1}(u_1) \cdots \chi_{\a_s}(u_s),
  \eeq
  {\it where $\chi_{\a_i}(u_i) = e^{2\pi i \a_i m_i}$ with integers $m_i$
 determined via $u_i = g^{m_i}$, where $g\in \mathF_q$ is a
generator. Then the cardinality of $X^{\un}/\mathF_q$ is given by}
 \beq
 \#(X^{\un}/\mathF_q) = N_{1,q}(X^{\un}) = 1+q + \cdots + q^{s-1}+
  \sum_{\a \in \cA_s^{q,\un}} j_q(\a).
 \eeq

With these ingredients the L-function associated to
$H^{s-1}(X^{\un})$ takes the form (\ref{lfunction}) with
polynomials $\cP_p^{s-1}(t)$ expressed in terms of the Jacobi-sums
variables by
 \beq
  \cP_p^{s-1}(t) = \prod_{\a \in \cA_s^{\un}}
   \left(1 - (-1)^{s-1} j_{p^f}(\a) t^f \right)^{1/f}
   \eeq
  where
\beq \cA_s^{\un} = \left\{\a \in \mathQ^{s+1} ~{\Big |}~
0<\a_i<1,~n_i\a_i =0~(\rmmod~1),~\sum_{i=0}^s \a_i=0~(\rmmod~1)
\right\}
 \eeq
 and $f$ is determined via
 \beq
 (p^f-1)\a_i =0~(\rmmod ~1),~~~\forall i.
  \eeq
 The main point now is to identify an appropriate factor of the L-function of
 $H^{s-1}(X^{\un})$.

\subsection{Galois representations and $\Om-$motives of Calabi-Yau varieties}

This section defines the notion of $\Om-$motives for the general
class of Calabi-Yau hypersurfaces. The Jacobi sums $j_{p^f}(\a)$
are algebraic numbers in the cyclotomic field $\mathQ(\mu_d)$,
generated by primitive $d^{\rmth}$ roots of unity. The factors
$(1-j_{p^f}(\a)t^f)^{1/f}$ therefore do not themselves define
string theoretic L-series of the type considered in \cite{su02,
ls04, s05} because the latter have integral coefficients. The
simplest way to produce real coefficients in the L$-$function is
to combine 'dual' pairs of Jacobi sums, i.e. sums parametrized by
$(\a,\a')$ such that $\a+\a'=1$. This sometimes leads to success,
but in general the L-function of such pairs has coefficients that
are elements of the maximal real subfield of the cyclotomic field.
The idea in this paper is to achieve the necessary integrality of
the coefficients by considering orbits of Jacobi sums defined by
the action of the multiplicative group
$(\mathZ/d\mathZ)^{\times}$, where $d=k_in_i$, $\forall i$, is the
degree of the hypersurface.  The fact that these orbits lead to
integral coefficients in the L$-$function can be derived from a
number theoretic result about finite extensions of number fields
by noting that the multiplicative groups
$(\mathZ/d\mathZ)^{\times}$ can be interpreted as Galois groups of
 cyclotomic fields $\mathQ(\mu_d)$. This can be seen as follows
(the relevant number theoretic concepts can be found in \cite{n99}
and \cite{n04}).

Jacobi sum orbits for the Galois group
 $\rmGal(\mathQ(\mu_d)/\mathQ) = (\mathZ/d\mathZ)^{\times}$ are
obtained by defining an action
 \beq
 (\mathZ/d\mathZ)^{\times} \times \cA_s^{\un} \lra \cA_s^{\un}
 \eeq
 as
  \beq
   (t,\a) = t\cdot \a(\rmmod ~1),
  \eeq
  where $t\cdot \a = (t\a_0,...,t\a_s)$. For any $\a \in \cA_s^{\un}$ one
  can therefore consider its orbit
  \beq
   \cO_{\a} = \left\{\b \in \cA_s^{\un}~|~\b = t\cdot \a,~t\in
   (\mathZ/d\mathZ)^{\times} \right\},
   \eeq
   and decompose the set $\cA_s^{\un}$ into a set of orbits
   \beq
    \cA_s^{\un} = \bigcup_{\a} \cO_{\a},
   \eeq
   where $\a$ is a representative in each orbit.

To see that Galois orbits lead to $L-$series with rationally
integral coefficients one observes that the Jacobi sums $j_p(\a)$
transform under the action of $t \in (\mathZ/d\mathZ)^{\times}$ as
 \beq
  j_p(t\cdot \a) = \si_t(j_p(\a)),
  \eeq
  where $\si_t \in \rmGal(\mathQ(\mu_d)/\mathQ)$ is defined as
  $\si_t(\xi_d) = \xi_d^t$. Hence the Galois orbit leads to
  coefficients $a_p$ of the terms with prime exponents that are of
  the form
  \beq
  a_p(\a) = \sum_{\si \in \rmGal(\mathQ(\mu_d)/\mathQ)} \si(j_p(\a)).
  \eeq

The argument can now be completed by noting that the sum $a_p(\a)$
is the trace of an algebraic integer and by using a result from
the theory of finite extensions.
 For any element $x\in L$ in a finite extension $L/K$
over a number field $K$ Dedekind defined the map
 \bea
  \rmT_x:  L ~&\lra& ~L \nn \\
           y ~&\mapsto &~ xy,
 \eea
 which can be used to define a trace map
  \beq \rmTr_{L/K}(x) := \rmtr ~\rmT_x. \eeq
 It turns out that this map takes values in the base field $K$, and that
 when restricted to the ring $\cO_L$ of algebraic integers of $L$
 it maps integers to integers
  \beq
  \rmTr_{L/K}: \cO_L  \lra \cO_K.
  \eeq
 The link to the above considerations of Jacobi sums is made by
 noting that for separable extensions $L/K$ one can show that
  \beq
   \rmTr_{L/K}(x) = \sum_{\si: L\ra \oK} \si(x)
  \eeq
  where $\oK$ denotes the closure of $K$.
 The cyclotomic fields of interest here are finite extensions
 of the rational field $\mathQ$, which is a perfect field,
 hence all its finite extensions are separable.
 This means that the trace maps the ring of
 algebraic integers of any cyclotomic field into the rational
 integers. It follows that the coefficients $a_p$ in the L$-$series are
 elements in $\mathZ$. Cyclotomic fields are normal, therefore the
 embeddings define the Galois group of automorphisms which leave
 $\mathQ$ invariant.

The action defined above induces an action on the middle
cohomology of any diagonal hypersurface $X^d$ of degree $d$
 embedded in a weighted projective space $\mathP_{(k_0,...,k_s)}$
with weights $k_i$ because the set $\cA_s^{\un}$ parametrizes the
(untwisted) cohomology classes in $H^{s-1}(X^d)$. The strategy to
compute the L$-$function of the orbits $\cO_{\a}$ determined by
the factorization
 \beq
  \cP_p^{s-1}(t) = \prod_{\cO_{\a}} \cP_p^{s-1}(\cO_{\a}, t),
  \lleq{decomposition}
 where
 \beq
  \cP_p^{s-1}(\cO_{\a},t)
  = \prod_{\beta \in \cO_{\a}} \left(1-j_{p^f}(\beta)t^f\right)^{1/f},
  \eeq
 therefore implies the computation of the L-function of the irreducible
representations of the Galois group
 of the cyclotomic field associated to the variety on the
 intermediate cohomology. From now on the superscript of the
 polynomial $\cP_p^{s-1}(t)$, indicating the cohomology group, will be
 dropped.

Of particular importance for Calabi-Yau spaces is the orbit
$\cO_{\Om}$ of the element $\a_{\Om}:= \left( {\footnotesize
\frac{k_0}{d},...,\frac{k_s}{d}}\right)$
 associated to the holomorphic $(s-1)-$form. This motivates the
 introduction of the following two definitions.

 {\bf Definition.}~{\it The Galois orbit $\cO_{\Om}$ of the element
 $\a_{\Om} \in \cA_s^{\un}$ with respect to the action of the Galois group
 $\rmGal(\mathQ(\mu_d)/\mathQ)$ is called the $\Om-$representation
 of the Galois group $\rmGal(\mathQ(\mu_d)/\mathQ)$.}

It is possible to associate to an $\Om-$representation defined by
$\cO_{\Om}$ a projection operator which can be used to define a
motive in the sense of Shioda.

 {\bf Definition.}~{\it The motive associated to the
 $\Om-$representation defined by the orbit $\cO_{\Om}$ will be called
 the $\Om-$motive $M_{\Om}$ of the variety.}

In this paper the focus will be on the computation of the
L-function $L_{\Om}(X,s):=L(M_{\Om},s)$ of the $\Om-$motive of
hypersurfaces $X^d$
 \beq
  L_{\Om}(X^d,s)
  = \prod_p \prod_{\a \in \cO_{\Om}}
   \frac{1}{\left(1-(-1)^{s-1}j_{p^f}(\a)p^{-fs}\right)^{1/f}},
  \eeq
where $\cO_{\Om} \subset \cA_s^{\un}$ denotes the orbit of the
element $\a$ that corresponds to the $\Om-$form. More precisely,
the aim is to check whether the $\Om-$motives of extremal
Brieskorn-Pham K3 surfaces admit a modular interpretation and if
so, whether the resulting forms admit an affine Lie algebraic
construction. Higher-dimensional varieties are considered in
\cite{s06b}.

\vskip .1truein

\section{Examples}

The general framework formulated above is applied in this paper to
extremal K3 surfaces of Brieskorn-Pham type. A K3 surface $S$
defined over $\mathC$ is called extremal if its Picard number
$\rho(S)=\rmrk~\rmNS(S)$, defined as the rank of the
N\'eron-Severi group $\rmNS(S)$, is maximal, i.e. $\rho(S)=20$.
Such surfaces were originally called singular \cite{si77}, and
more recently have been called attractive \cite{m98}. The set of
extremal Brieskorn-Pham K3 surfaces is given in eq.
(\ref{examples}).

\subsection{The quartic Fermat K3 surface $S^4$}

A summary of cardinality results $N_{r,p}(S^4)$ for small primes
$p$ for the Fermat K3 surface of degree four is contained in Table
1.

\begin{center}
\begin{tabular}{l| r r r r r r r r r r r r}
Prime $p$     &3     &5     &7     &11    &13    &17
              &19    &23    &29    &31    &37    \tabroom \\

\hline

$N_{1,p}(S^4)$ &16    &0     &64    &144   &128    &600
              &400   &576   &768   &1024  &1152   \tabroom \\

\hline

$\beta_1(p)$  &$-6$  &0     &$-14$ &$-22$ &42 &$-310$
              &$-38$ &$-46$ &74    &$-62$ &218     \tabroom \\
\hline
\end{tabular}
\end{center}

{\bf Table 1.}{\it ~The coefficients
$\beta_1(p)=1+p^2-N_{1,p}(S^4)$ of the Hasse-Weil modular form of
the quartic Fermat surface $S^4$ in terms of the cardinalities
$N_{1,p}$ for small primes.}

The results in Table 1 lead to the L-series of $S^4$
 \beq
   L(S^4,s)  \doteq 1 + \frac{6}{3^s}  - \frac{26}{5^s} +
\frac{14}{7^s} + \frac{117}{9^s} + \frac{22}{11^s} -
\frac{42}{13^s} - \frac{156}{15^s} + \frac{310}{17^s} +
\frac{38}{19^s} + \frac{84}{21^s} + \frac{46}{23^s} +, \cdots
 \eeq
where the symbol $\doteq$ means that a finite number of Euler
factors have been omitted, here the bad prime $p=2$. Associated to
this L-series is the $q-$expansion
 \beq
 f(S^4,q) \doteq q + 6q^3 - 26q^5 + 14q^7 + 117q^9 +
  22q^{11} - 42q^{13} -156q^{15} + 310q^{17} + 38q^{19} + 84q^{21} + 46q^{23} + \cdots
 \eeq

The product structure of $L(S^4,s)$ can be obtained from
 the Jacobi-sum formulation by enumerating
the set $\cA^{\un}_3$ of the surface $S^4$. Replacing the degree
vector $\un$ by the degree itself leads to
 \bea
 \cA^4_3 &=&
 \left\{
 \left({\tiny \frac{1}{4}}, \frac{1}{4}, \frac{1}{4}, \frac{1}{4}\right),~
 \left({\small \frac{1}{2}, \frac{1}{2}, \frac{1}{2}, \frac{1}{2}}\right),~
 \left({\small \frac{3}{4}, \frac{3}{4}, \frac{3}{4}, \frac{3}{4}}\right),~
 \left({\small \frac{1}{4}, \frac{1}{2}, \frac{1}{2}, \frac{3}{4}}\right),~
 \left({\small \frac{1}{4}, \frac{1}{4}, \frac{3}{4},
 \frac{3}{4}}\right)\right\} \nn \\
 & & ~~~+ {\rm permutations}.
 \eea

The Jacobi sums of the K3 surface $S^4$ at low primes are
collected in Table 2. In this table the permutations and the
complex conjugates of the sums listed are suppressed.

\begin{small}
\begin{center}
\begin{tabular}{l| c| c c c c c c }
Type  &$q$            &5      &9     &13         &17        &29
               &37    \tabroom \\
\hline

I   &$j_q\left({\small \frac{1}{4}, \frac{1}{4}, \frac{1}{4},
\frac{1}{4}} \right)$
           &$-3+4i$    &9     &$5-12i$  &$-15-8i$   &$21+20i$
           &$-35-12i$  \tabroom \\

II  &$j_q\left({\footnotesize \frac{1}{2},
\frac{1}{2},\frac{1}{2},\frac{1}{2}}\right)$
               &\phantom{+}5       &9     &\phantom{+}13   &17
               &\phantom{+}29      &\phantom{+}37  \tabroom \\

 III &$j_q\left({\small \frac{1}{4}, \frac{1}{4}, \frac{3}{4}, \frac{3}{4}}\right)$
                &\phantom{+}5      &9     &\phantom{+}13  &17
                &\phantom{+}29     &\phantom{+}37  \tabroom \\

IV  &$j_q\left({\small \frac{3}{4}, \frac{1}{4}, \frac{1}{2},
\frac{1}{2}}\right)$
               &$-5$    &9    &$-13$   &17  &$-29$ &$-37$  \tabroom \\
 \hline \hline
 \end{tabular}
\end{center}
\end{small}

\centerline{{\bf Table 2.}{\it ~Jacobi-sums of the quartic K3
surface $S^4 \subset \IP_3$.}}

The L-function of $S^4$ therefore factorizes as
 \beq
  L(S^4,s) = L_{\rm I}(S^4,s) \cdot L_{\rm II}(S^4,s) \cdot
                     L_{\rm III}(S^4,s) \cdot L_{\rm IV}(S^4,s),
 \eeq
 where the individual factors correspond to the orbits of the different
 Jacobi sums of Table 2. The first factor describes the L-function
 of the $\rmGal(\mathQ(i)/\mathQ)-$orbit of
 $j_p\left(\footnotesize{\frac{1}{4},\frac{1}{4}, \frac{1}{4},
 \frac{1}{4}}\right)$, which corresponds to the holomorphic
 2-form of $S^4$, and hence is the $\Om-$motivic L-function of $S^4$,
 $L_{\Om}(S^4,s)=L_{\rm I}(S^4,s)$ of $S^4$,
 \bea
  L_{\Om}(S^4,s) &=&
  \prod_{p\neq 2}
  \frac{1}{\left(1-j_{p^f} \left({\tiny \frac{1}{4},
      \frac{1}{4},\frac{1}{4},\frac{1}{4}}\right) \cdot
      p^{-fs}\right)^{1/f}
      \left(1-j_{p^f}\left({\tiny \frac{3}{4},
      \frac{3}{4},\frac{3}{4},\frac{3}{4}}\right) \cdot p^{-fs}\right)^{1/f}}
      \cdot \nn \\
  & & \nn \\
 &\doteq & 1-\frac{6}{5^s} + \frac{9}{9^s} + \frac{10}{13^s} - \frac{30}{17^s}
    + \frac{11}{25^s} + \frac{42}{29^s} - \frac{70}{37^s} + \cdots
 \eea
 The factor $L_{\Om}(S^4,s)$ of the complete L-function of $S^4$
 is the only one with Jacobi sum characters
 in an algebraic number field.

\subsection{The degree six weighted K3 surface $S^{6\rmA}$}

The L-series of the $\Om-$motive $M_{\Om}$ of the double cover
$S^{6\rmA}$ of the projective plane branched over a degree six
plane curve  is determined by the Jacobi sums
 \beq
  j_p{\small \left(\frac{1}{6},\frac{1}{6},\frac{1}{6},\frac{1}{2}
      \right)},~~~~~~
   j_p{\small \left(\frac{5}{6},\frac{5}{6},\frac{5}{6},\frac{1}{2}
      \right)},
  \eeq
 which are complex conjugates. The values for these sums
 for low primes are collected in Table 3.

 \begin{center}
 \begin{tabular}{l| c c c c c }
 Prime $p$                &7    &13    &19   &31   &37   \tabroom  \\

\hline

 $j_p{\footnotesize \left(\frac{1}{6},\frac{1}{6},\frac{1}{6},\frac{1}{2}
      \right)}$   &$-\frac{13}{2} - \frac{3}{2}\sqrt{3}i$
                   &$-\frac{1}{2} + \frac{15}{2}\sqrt{3}i$
                   &$\frac{11}{2} - \frac{21}{2}\sqrt{3}i$
                   &$-23+12\sqrt{3}i$
                   &$\frac{47}{2} - \frac{33}{2}\sqrt{3}i$
\tabroom \\
\hline
\end{tabular}
\end{center}

\centerline{{\bf Table 3.}~{\it Jacobi sums for the K3 surface
$S^{6\rmA}\subset \mathP_{(1,1,1,3)}$ for small primes.}}

The resulting L-series of the $\Om-$orbit is given by
 \beq
  L_{\Om}(S^{6\rmA},s) \doteq
  1 - \frac{13}{7^s} - \frac{1}{13^s} + \frac{11}{19^s}
  - \frac{46}{31^s} + \frac{47}{37^s} + \cdots
 \lleq{ls6}
 up to a finite number of Euler factors.

\subsection{The degree six weighted K3 surface $S^{6\rmB}$}

The $\Om-$motivic L-series of the degree six Brieskorn-Pham type
surface embedded in $\mathP_{(1,1,2,2)}$ can be computed via the
Jacobi sums
  \beq
  j_p{\small \left(\frac{1}{6},\frac{1}{6},\frac{1}{3},\frac{1}{3}
      \right)},~~~~~~
   j_p{\small \left(\frac{5}{6},\frac{5}{6},\frac{2}{3},\frac{2}{3}
      \right)}
  \eeq
 whose values at low primes are collected in Table 4.

\begin{small}
\begin{center}
\begin{tabular}{l| c c c c c}
$p$          &7                     &13
             &19                    &31
             &37   \tabroom \\
\hline

 $j_p\left({\small \frac{1}{6}, \frac{1}{6}, \frac{1}{3}, \frac{1}{3}}\right)$
             &$-1-4\sqrt{-3}$       &$-11-4\sqrt{-3}$
             &$-13-8\sqrt{-3}$      &$23-12\sqrt{-3}$   &$13+20\sqrt{-3}$
      \tabroom \\
 \hline
\end{tabular}
\end{center}
\end{small}

\centerline{{\bf Table 4.}{\it ~~Jacobi-sums for the K3 surface
$S^{6\rmB} \subset \mathP_{(1,1,2,2)}$.}}

The orbit of the $\Om-$form with respect to the Galois group
$\mathQ(\mu_6)/\mathQ$ leads to the L-series
 \beq
  L_{\Om}(S^{6\rmB},s) \doteq 1 - \frac{2}{7^s} - \frac{22}{13^s} -
      \frac{26}{19^s} + \frac{46}{31^s} + \frac{26}{37^s} + \frac{22}{43^s}
       +  \cdots
  \eeq

The question that arises now is whether these expansions can be
given a string theoretic meaning, similar to the case of elliptic
Brieskorn-Pham curves.

\vskip .1truein

\section{CFT Modularity of $\Om-$motives}

\subsection{Modular forms}

One of the important ingredients in the analysis of modular
L-functions is their product structure, hence it is of interest to
consider forms which admit Euler products, i.e. Hecke eigenforms.

{\bf Definition.}
 {\it A modular form of weight $w$, level $N$, and character $\chi$
  with respect to $\Gamma_0(N)$ is a map $f: \cH \lra \mathC$ on the upper
  half$-$plane such that for any $\tau \in \cH$ and
  $\g = {\scriptsize \left(\matrix{a&b\cr c&d\cr}\right)} \in \Gamma_0(N)$}
  \beq
  f(\g \tau) = \chi(d) (c\tau +d)^w f(\tau).
 \eeq

A cusp form can be characterized by the condition that the general
$q-$expansion $f(\tau) = \sum_{n=0}^{\infty} a_nq^n$ starts with
$a_1$. It is normalized if $a_1=1$.

Hecke defined on the space of cusp forms $S_w(\Gamma_0(N),\chi)$
 operators which for prime $p$ take the form
  \beq
   T_p^wf(q) = \sum_{n=1}^{\infty} a_{np}q^n + \chi(p)p^{w-1}
   \sum_{n=1}^{\infty} a_nq^{np}.
  \eeq
 Eigenforms of these operators are particularly important in a
 geometric context.

{\bf Theorem 4.}(Hecke) \hfill \break
 {\it The space
$S_w(\Gamma_0(N),\chi)$ of cusp forms of weight $w$, level $N$,
and character $\chi$ is the orthogonal sum of the spaces of
equivalent eigenforms. Each space of such forms has a member that
is an eigenvector of all Hecke operators $T_w(n)$. A form $f \in
S_w(\Gamma_0(N), \chi)$ that is an eigenvector for all $T_w(n)$
can be normalized and its coefficients satisfy}
 \bea
 a_{p^{n+1}} &=& a_{p^n}a_p - \chi(p) p^{w-1}a_{p^{n-1}},~~~~~
                p \notdiv N, \nn  \\
 a_{p^n} &=& (a_p)^n,~~~~~p|N, \nn\\
 a_{mn}  &=& a_ma_n,~~~~~(m,n)=1.
 \eea
 {\it Moreover, the L-function $L(f,s)$ has an Euler product of the form}
 \beq
 L(f,s) = \prod_{\stackrel{p~{\rm prime}}{p|N}} \frac{1}{1-a_pp^{-s}}
       \prod_{\stackrel{p~{\rm prime}}{p\notdiv N}}
        \frac{1}{1-a_pp^{-s}+p^{w-1-2s}}
 \eeq
 {\it which is convergent for $\rmRe~s>\frac{w}{2}+1$.}

\subsection{CFT ingredients}

Quantities that have proven useful in the string modularity
analysis of elliptic Brieskorn-Pham curves \cite{su02, ls04, s05}
are the theta functions
 \beq
 \Theta^k_{\ell, m}(\tau)
 = \sum_{\stackrel{\stackrel{-|x|<y\leq |x|}{(x,y)~{\rm or}~
      (\frac{1}{2}-x,\frac{1}{2}+y)}}{\in
\mathZ^2+\left(\frac{\ell+1}{2(k+2)},\frac{m}{2k}\right)}}
\rmsign(x) e^{2\pi i \tau((k+2)x^2-ky^2)},
 \eeq
 defined by Kac and Peterson. These are related to the string functions
 of the affine algebra $A_1^{(1)}$ at level $k$ as
 \beq
 c^k_{\ell,m}(\tau) = \frac{\Theta^k_{\ell,m}(\tau)}{\eta^3(\tau)}.
 \eeq
 The string functions, together with the classical theta functions
 \beq
 \theta_{n,m}(\tau,z,u) =
e^{-2\pi i m u} \sum_{\ell \in \mathZ + \frac{n}{2m}} e^{2\pi i m
\ell^2 \tau + 2\pi i \ell z},
 \eeq
 are the building blocks of the N$=$2 supersymmetric characters
 $\chi^k_{\ell,q,s}(\tau)$ of the partition functions of the $N=2$ minimal theories
 \bea
  \chi^{k}_{\ell,q,s}(\tau,z,u) &=&
   e^{2\pi i u} \rmtr_{\cH^{k}_{\ell,q,s}}
    q^{\left(L_0 - \frac{c}{24} \right)} e^{2\pi i zJ_0} \nn \\
  &=& e^{2\pi i u} \sum_{Q^{k}_{\ell,q,s}, \Delta^{k}_{\ell,q,s}}
   {\rm mult}\left(\Delta^{k}_{\ell,q,s}, Q^{k}_{\ell,q,s}\right)
   e^{2\pi i\tau \left(\Delta^{k}_{\ell,q,s} -
          \frac{c}{24}\right)+ 2\pi i zQ^{k}_{\ell,q,s}},
 \eea
 in terms of the conformal dimensions $\Delta^k_{\ell,q,s}$ and
 the charges $Q^k_{\ell,q,s}$, leading to \cite{g88}
\begin{equation}
 \chi^{k}_{\ell,q,s}(\tau, z, u) = \sum
c^{k}_{\ell,q+4j-s}(\tau) \theta_{2q+(4j-s)(k+2),2k(k+2)}(\tau,
z,u).
\end{equation}

It will become clear below that the modular forms that explain the
modularity of extremal K3 surfaces of Brieskorn-Pham type in a
string theoretic way are given by
 \bea
  \Theta^1_{1,1}(q) &=& q^{1/12}(1 - 2q -q^2 + 2q^3 + q^4 + 2q^5 -
     2q^6 - 2q^8 - 2q^9 + q^{10} + \cdots ) \nn \\
  \Theta^2_{1,1}(q) &=& q^{1/8}(1 - q - 2q^2 + q^3 + 2q^5 + q^6 - 2q^9
   + q^{10} + \cdots ).
 \eea

\subsection{Modularity of $\Om-$motives}

If an $\Om-$motive of K3 type is modular the associated modular
form is expected to be of weight three. A useful guide in the
search for such forms is provided by the speculation that exactly
solvable models lead to motives which admit complex multiplication
(CM) in the classical sense \cite{lss04} (see \cite{y06} for an
alternate discussion of CM). The quartic Fermat surface admits CM
with respect to $\mathQ(\sqrt{-1})$, while the degree six surfaces
have CM with respect to the field $\mathQ(\sqrt{-3})$. The goal
therefore is to find modular forms of weight three which admit
complex multiplication with respect to these fields.

A further guide is provided by the bad primes of these surfaces.
It is expected that the level of a geometrically derived modular
form is divisible by the bad primes of the underlying variety. The
only bad prime for the quartic Fermat K3 surface is $p=2$, hence
the level should be some power of two. For the degree six surfaces
the bad primes are $p=2,3$, hence the level should be of the form
$2^a3^b$ for some non-negative integers $a,b$.

Combining these considerations leads to the candidate cusp forms
 \bea
 S_3(\Gamma_0(16),\chi_{-1}) &\ni & \eta^6(4\tau)
 = q-6q^5+9q^9 +10q^{13}-30q^{17} + 11q^{25} + \cdots \nn \\
 S_3(\Gamma_0(12),\chi_{-3}) &\ni &
 \eta^3(2\tau)\eta^3(6\tau) = q - 3q^3 + 2q^7 +9q^9 -22q^{13} +
 26q^{19}  + \cdots
 \eea
  for the K3 surfaces considered above.
 Comparing the $q-$expansions of these two forms with those of the
 $\Om-$motivic L-series of the K3 surfaces
 shows that while the first form describes the L$-$function of the
 quartic K3 surface, the latter describes the corresponding L$-$function
 of $S^{6\rmB}$ only up to sign changes. This indicates
that a twist is involved, and it turns out that this twist can be
provided by the Legendre character $\chi_3$. This leaves the
surface $S^{6\rmA}$. A candidate modular form can be obtained by
lifting a complex multiplication modular form of weight two to
weight three via an Eisenstein series associated to the CM field
$\mathQ(\sqrt{-3})$. In the present case a useful modular form
turns out to come from a class of theta series considered in
\cite{h25}. Hecke associates to each element $\a$ of an integral
ideal $\afrak$ in an imaginary
 quadratic field $K=\mathQ(\sqrt{-D})$ the theta series defined as
 \beq
  \vartheta(\tau; \a, \afrak, Q\sqrt{-D}) =
   \sum_{z\equiv \a(\rmmod~\afrak Q \sqrt{-D})} q^{\rmN
   z/Q|D|\rmN\afrak},
  \eeq
  where $\rmN z$ and $\rmN\afrak$ denote the norms of
   $z\in \cO_K$ and $\afrak$ respectively, and $\cO_K$ is the ring of
  integers of $K$.
 Relevant for the L-series $L_{\Om}(S^{6\rmA},s)$ is the special
 case given by $\a=0$, $Q=1$, and $\afrak = \cO_K$ for the
 Eisenstein field $K=\mathQ(\sqrt{-3})$, renamed here as
 \beq
  \vartheta(q) = \sum_{z\in \cO_K} q^{\rmN z}.
 \eeq
In summary, the results above lead to the identification of the
respective Mellin transforms $f_{\Om}(S^d,q)$ of the motivic
L$-$functions $L_{\Om}(S^d,s)$ for $d=4,6\rmA,6\rmB$ as noted in
eq. (\ref{k3mod}) of Theorem 1.
 The proof of these relations to all orders is postponed to the appendix.

The modularity of the $\Om-$motives of extremal Brieskorn-Pham K3s
leaves the question whether the resulting modular forms admit a
string theoretic interpretation in terms of forms derived from the
associated conformal field theory. The basic idea in the following
is to consider the arithmetic building blocks of the K3 surfaces
considered here. One way to identify these structures is via the
twist map. In this section the focus will remain on the arithmetic
aspects, postponing the identification of the basic irreducible
geometric structure to section 5.

\subsection{Elliptic curves from K3 surfaces}

The structure of the quartic K3 surface $S^4$ is very simple. It
is shown below that it can be constructed directly in terms of
$E^4$ via the twist map \cite{hs99}. Alternatively, one can use
the Shioda-Katsura decomposition \cite{sk79, d82} to reconstruct
the cohomology of $S^4$ from the cohomology of the quartic plane
curve
 \beq C_4 =
 \{(z_0:z_1:z_2)\in \mathP_2~|~ z_0^4 + z_1^4 +z_2^4 =0 \}.
 \eeq
 This is a genus three curve whose L-function factors into the
 triple product of the L-function of the weighted elliptic curve
 $E^4$.
The L-function of this curve was computed in \cite{ls04}, where it
was shown that its Mellin transform is determined by a twist of
the Hecke indefinite modular form $\Theta^2_{1,1}(q)$, as
described in Theorem 2. Therefore the geometry and arithmetic
structure of the quartic Fermat K3 suggests a relation between the
modular forms $\eta^6(q)$ and $\Theta^2_{1,1}(q)$ of weight three
and two, respectively.

The structure of the degree six K3 surface $S^{6\rmB}$ can be
recovered from the twist map in a way similar to the quartic
surface by constructing it from two copies of the elliptic curve
$E^6$, as described below.  Alternatively, for both degree six
surfaces one can again consider the reduction of the cohomology
via Shioda-Katsura. There are two curves to consider, the weighted
plane curve
 \beq
  C_6 = \{(z_0:z_1:z_2)\in \mathP_{(1,1,2)}~|~z_0^6+z_1^6+z_2^3=0
  \}
 \eeq
 and the elliptic curve $E^3$. The latter was analyzed in detail in
 \cite{su02}, with the result that its associated modular form is determined
 by the theta series $\Theta^1_{1,1}(q)$ at conformal level $k=1$.
 The Jacobian of $C_6$ factors into three different types of elliptic curves,
 $E^3$ just discussed, the
 degree six elliptic curve $E^6$,
  and a third curve of conductor $432$. The curve $E^6$ has been
  shown to lead to the string theoretic modular form given in terms
  of $\Theta^1_{1,1}(q)$. These results then suggest a
  relation between the modular form $\eta^3(2\tau)\eta^3(6\tau)$
  and $\Theta^k_{1,1}(q)$ for levels $k=1$ or $k=2$, or both.

\subsection{From K3 to CFT forms}

For the degree six surface $S^{6\rmA}$ the relation between the
weight three form and the string theoretic form is immediate
because this form is the lift of a product of Hecke indefinite
modular forms. In general, the expected relation between forms of
different weight cannot be established for the forms themselves,
but should proceed in terms of their associated L-series. Guidance
for such constructions is provided by the theory of convolutions
of L-functions. It turns out that for all three modular forms of
weight three determined above the relation is of similar type.
Denote by
 \beq
 f_w(q) = \sum_{n=1}^{\infty} a^{(w)}_n q^n
 \eeq
 the $q-$expansion of
the weight $w$ form. Then the relation between the pairs of weight
three and weight two forms for the surfaces
 \bea
  S^4:~~(f_3,f_2) &=& (\eta^6(q^4),\Theta^2_{1,1}(q^4)^2\otimes \chi_2),
                               \nn \\
  S^{6\rmA}:~~(f_3,f_2) &=& (\vartheta(q^3)\eta^2(q^3)\eta^2(q^9),
                             \Theta^1_{1,1}(q^3)\Theta^1_{1,1}(q^9))
 \eea
 is of the form
  \beq
   a_p^{(3)} = (a_p^{(2)})^2 - 2p
  \lleq{3-2relation1}
  for rational primes $p$, while for third surface
  \beq
  S^{6\rmB}:~~(f_3,f_2) = (\eta^3(q^2)\eta^3(q^6),\Theta^1_{1,1}(q^6)^2\otimes \chi_3),
  \eeq
  a twist is necessary
   \beq
   a_p^{(3)} = \left((a_p^{(2)})^2 - 2p\right)\chi_3(p).
  \lleq{3-2relation2}
  The origin of these relations, and the motivation for the
  various twists, will become clear in the next two sections.

This result shows that the string theoretic modular forms
$\Theta^k_{\ell,m}(\tau)$ introduced above suffice to explain the
arithmetic structure of spacetime, represented here by the
$\Om-$motive carrying an irreducible representation of the Galois
group of the cyclotomic field, determined by the symmetry group of
the K3 surfaces. Reading these relations in reverse explains
string theoretic modular forms in terms of the arithmetic geometry
of the topologically nontrivial part of spacetime.

\vskip .1truein

\section{K3 surfaces from elliptic curves}

This section describes the elliptic building blocks used in the
previous section to identify the string theoretic structure of
extremal Brieskorn-Pham K3s. A direct construction of the two
surfaces $S^4$ and $S^{6\rmB}$ in terms of elliptic curves can be
obtained via the twist map \cite{hs99}. This construction was
originally considered in the context of string dualities
\cite{hs95} (see also the later ref. \cite{b97}), and will also be
useful below for the proof of modularity to all orders for the
$\Om-$motivic L-series of these K3 surfaces.

\subsection{The twist map construction}

 In the notation of \cite{hs99} consider the map
 \beq
 \Phi: ~\mathP_{(w_0,...,w_m)} \times \mathP_{(v_0,...,v_n)}
  \lra \mathP_{(v_0w_1,...,v_0w_m,w_0v_1,...,w_0v_n)}
 \eeq
 defined as
  \bea
  ((x_0,...,x_m),(y_0,...,y_n)) &\mapsto &
   (y_0^{w_1/w_0}x_1,...,y_0^{w_m/w_0}x_m,
   x_0^{v_1/v_0}y_1,...,x_0^{v_n/v_0}y_n) \nn \\
   ~~~& & =:(z_1,...,z_m,t_1,...,t_n).
   \eea
 This map restricts on the subvarieties
  \bea
   X_1 &=& \{x_0^{\ell} + p(x_i) =0\} \subset \mathP_{(w_0,...,w_m)}
   \nn \\
   X_2 &=& \{y_0^{\ell} + q(y_j) =0\} \subset
   \mathP_{(v_0,...,v_n)},
  \eea
  defined by transverse polynomials $p(x_i)$ and $q(y_j)$,
  to the hypersurface
  \beq
   X = \{p(z_i) - q(t_j) =0\} \subset
    \mathP_{(v_0w_1,...,v_0w_m,w_0v_1,...,w_0v_n)}
   \eeq
  as a finite map.
  The degrees of the hypersurfaces $X_i$ are given by
   \beq
    \rmdeg~X_1 = w_0\ell,~~~~~~\rmdeg~X_2 = v_0\ell,
   \eeq
   leading to the degree $\rmdeg~X = v_0w_0\ell$.

\subsection{Examples}

Applying the twist construction to the quartic elliptic curve $E^4
\subset \mathP_{(1,1,2)}$ leads first to the map of ambient spaces
  \beq
 \Phi:   \mathP_{(2,1,1)} \times \mathP_{(2,1,1)} \lra
   \mathP_3
  \eeq
 defined as
 \beq
  ((x_0,x_1,x_2),(y_0,y_1,y_2)) ~\mapsto ~
  (y_0^{1/2}x_1,y_0^{1/2}x_2,x_0^{1/2}y_1,x_0^{1/2}y_2).
  \eeq
  Restricting the map $\Phi$ to the product $E^4_{+}\times E^4_{-}$ of the
  elliptic curves
  \beq
   E^4_{\pm} = \{x_0^2\pm (x_1^4+x_2^4) = 0\} \subset
   \mathP_{(2,1,1)},
  \eeq
 leads to the quartic K3 surface $S^4\subset \mathP_3$.

The degree six surface $S^{6\rmB}$ can be constructed similarly by
considering the elliptic curve $E^6 \subset \mathP_{(1,2,3)}$. The
ambient space map
 \beq
 \Phi: \mathP_{(3,2,1)} \times \mathP_{(3,2,1)} \lra
   \mathP_{(1,1,2,2)}
  \eeq
  is now defined by
 \beq
  ((x_0,x_1,x_2),(y_0,y_1,y_2)) ~\mapsto ~
  (y_0^{2/3}x_1,y_0^{1/3}x_2, x_0^{2/3}y_1,x_0^{1/3}y_2),
 \eeq
 and restricts on the product $E^6_{+} \times E^6_{-}$,
 where
  \beq
   E^6_{\pm} = \{x_0^2 \pm (x_1^3 + x_2^6)=0\} \subset
   \mathP_{(3,2,1)},
  \eeq
  to the surface $S^{6\rmB} \subset \mathP_{(1,1,2,2)}$,

 It will become clear below that the surface $S^{6\rmA}$ is
 determined by the elliptic curve $E^3$. Therefore all surfaces
 considered here can be understood in terms of the arithmetic
 structure of the three Brieskorn-Pham type elliptic curves
 described in \cite{s05}.

\subsection{K3 geometry from string theory}

One of the fundamental problems in string theory is to derive the
structure of spacetime solely from the field theory on the
worldsheet. In refs. \cite{su02,s05} it was shown that it is
possible to construct the elliptic geometry of Gepner models at
$c=3$ directly from the conformal field theory, providing a string
theoretic derivation of the extra dimensional geometry for a
simple class of exact models. The question arises whether such a
derivation can be generalized to the K3 surfaces analyzed here. In
contrast to elliptic curves, the cohomology of K3 surfaces is more
complicated, and it is not obvious how a direct construction
should proceed. The twist map described above does, however, lead
directly from the modular forms on the worldsheet to the geometry
of spacetime in this higher dimensional case as well. This can be
seen as follows.

The first step is the construction, described above, of these K3
surfaces in terms of the elliptic curves of Brieskorn-Pham type
via the twist map. This reduces the modular construction from two
dimensions to one. The second step is to use the criteria
formulated in \cite{s05}, which uniquely identify the modular
forms for these elliptic curves. This leads to the modular forms
of weight 2 described in the previous section. This two-step
procedure can be reversed by using the Eichler-Shimura
construction, which allows to construct elliptic curves from their
modular forms. Combining these modularity arguments with the twist
map therefore leads to the construction of the K3 surfaces $S^4$
and $S^{6\rmB}$ directly from the conformal field theories on the
worldsheet. An alternative reduction strategy proceeds via complex
multiplication. This approach is less direct, but allows to
discuss all three surfaces in a unified manner, as shown below.

\vskip .1truein

\section{Modularity of a phase transition}

The arithmetic structure of a projective variety is sensitive to
deformations, changing as the coefficients of its defining
polynomials are varied. It is therefore of interest to consider
the behavior of families of varieties and analyze their arithmetic
structure as their moduli change. This line of thought has already
been followed in recent work \cite{cdr00, rv03, k04}. These
interesting papers address, among other issues,
 the question of what happens to the reduced variety at the
 conifold locus. Reference \cite{k04} in particular considers the
 lower dimensional analog of the conifold transition in the context
 of the following family of quartic K3 surfaces
 \beq
   S^4(\psi) = \left\{(z_0:\cdots :z_3)\in \mathP_3~{\Big
   |}~\sum_{i=0}^3 z_i^4 - 4\psi \prod_i z_i=0 \right\}.
 \eeq
 The result is that at the singular locus $\psi=1$ the congruent
 zeta functions $Z(S^4(\psi)/\mathF_p,t)$ degenerate,
 as expected. It turns out that they become singular in a way
 that preserves modularity. The question therefore arises whether
 there is a cohomological L-function whose Mellin
 transform admits a conformal field theoretic interpretation.

 The zeta function computations of ref. \cite{k04} indicate that
 at the singular point a modular form emerges which can be given
 a string theoretic interpretation as
 \beq
 \frac{(\Theta^2_{1,1}(\tau) \Theta^2_{1,1}(4\tau))^2}{\Theta^2_{1,1}(2\tau)}
 \otimes \chi_2.
  \eeq
 What is interesting about this form is that it is written
 in terms of cusp forms determined by precisely the same
  conformal field theory
 that leads to the L-function at the Fermat locus in the moduli
 space. Furthermore, the twist character
  $\chi_2$ which appears in this expression,
  is the Legendre character of the field of quantum
  dimensions $\mathQ(\sqrt{2})$ of that same theory.

  This suggests that at the analog of the conifold point the arithmetic
  structure points to a conformal field theory at level
  $k=2$, precisely the same structure that describes
  the situation for the Fermat surface.

\section{Symmetries}

\subsection{Complex multiplication}

It was conjectured in \cite{lss04} that exactly solvable
Calabi-Yau varieties can be characterized by a complex
multiplication symmetry in the sense of \cite{lps03}. In the
context of elliptic curves this coincides with the classical
notion of CM, and it was shown in \cite{s05} that all elliptic
Brieskorn-Pham curves and their associated modular forms admit CM
by either the Eisenstein field $\mathQ(\sqrt{-3})$ or the Gauss
field $\mathQ(\sqrt{-1})$. For higher dimensional Calabi-Yau
varieties the concept of complex multiplication must be modified.
The idea of ref. \cite{lps03} is to define the CM property of a
general Calabi-Yau variety in terms of the complex multiplication
properties of the associated Jacobians of the curves embedded in
the variety. This notion can be formulated more generally in terms
of the CM properties of motives defined by the Galois
representations, as described above.

A natural problem therefore is to determine the CM nature of the
forms derived above from the $\Om-$motives of extremal
Brieskorn-Pham K3 surfaces. In terms of the coefficients of the
$q-$expansion, $f(q)=\sum_n a_nq^n$, CM implies the existence of
quadratic extension fields $K/\mathQ$ such that the coefficients
$a_p$ vanish at all primes $p$ that are inert in $K$. In
\cite{s05} this condition was checked for the modular forms
associated to elliptic Brieskorn-Pham curves. Similarly one can
show that the forms of weight 3 of Theorem 1 admit CM. Table 5
summarizes the results for the CM fields of all the forms
encountered so far. The results indicate that the relevance of
complex multiplication generalizes from exactly solvable elliptic
curves to higher dimensions.

\begin{center}
\begin{tabular}{l | c c c | r }

Form                   &Weight  &Level   &CM Field
                       &Geometry\tabroom \\
     & & & & \\
\hline

$\Theta^1_{1,1}(q^3)\Theta^1_{1,1}(q^9)$
                       &2       &27      &$\mathQ(\sqrt{-3})$
                       &Elliptic curve  \\
  &  &  &  &  \\

$(\Theta^1_{1,1}(q^6))^2$
                       &2       &36      &$\mathQ(\sqrt{-3})$
                       &Elliptic curve   \\
  &  &  &  &  \\

$(\Theta^2_{1,1}(q^4))^2\otimes \chi_2$
                       &2       &64     &$\mathQ(\sqrt{-1})$
                       &Elliptic curve  \\
\hline

$\eta^6(q^4)$          &3       &16     &$\mathQ(\sqrt{-1})$
                       &K3 $\Om-$motive \\
&  &  &  &  \\

$\vartheta(q^3)\eta^2(q^3)\eta^2(q^9)$
                       &3        &27    &$\mathQ(\sqrt{-3})$
                       &K3 $\Om-$motive  \\
 & & & & \\

$\eta^3(q^2)\eta^3(q^6)\otimes \chi_3$
                       &3       &48     &$\mathQ(\sqrt{-3})$
                       &K3 $\Om-$motive  \\
                      &  &  &  &  \\
\hline
$\frac{\left(\Theta^2_{1,1}(q)\Theta^2_{1,1}(q^4)\right)^2}{\Theta^2_{1,1}(q^2)}$
                       &3       &8      &$\mathQ(\sqrt{-2})$
                       &Singular quartic K3 \tabroom \\
\hline
\end{tabular}
\end{center}

\centerline{{\bf Table 5.}~{\it CM modular forms with geometric
interpretation.}}

The complex multiplication property of the K3 motivic modular
forms provides an alternative approach to the notion of
"dimensional reduction" of K3 surfaces to elliptic curves
discussed in $\S 4$ and $\S 5$. CM modular forms can be obtained
from L-series
 \beq
  L(\psi,s) = \prod_{\pfrak \in \rmSpec~\cO_K}
   \frac{1}{1-\frac{\psi(\pfrak)}{\rmN\pfrak^s}}
   = \sum_{\afrak \subset \cI(O_K)}
    \frac{\psi(\afrak)}{\rmN\afrak^s}
  \eeq
  associated to Hecke characters $\psi$ of number fields $K$,
  where $\cI(\cO_K)$ ($\rmSpec~\cO_K$) describes the set of (prime) ideals in
  the ring of algebraic integers $\cO_K$, and $\rmN\afrak$ is the norm of the
  ideal $\afrak$ \cite{r77}. In the present case a Hecke
  characters $\psi_4$ of $\mathQ(\sqrt{-1})$, and two characters
  $\psi_{6\rmA}$ and $\psi_{6\rmB}$ associated to
  $\mathQ(\sqrt{-3})$, all of weight two,
   determine the L-series of $S^4$, $S^{6\rmA}$ and $S^{6\rmB}$ by
   considering the square of the basic characters.
 More precisely, the arithmetic of the $\Om-$motive of elliptic
 Brieskorn-Pham K3s can be obtained
 as follows. Define the character $\psi_4$ at the prime ideals $\pfrak$ of
 $\mathQ(\sqrt{-1})$ dividing the rational prime $p$ as
  \beq
   \psi_4(\pfrak) = \a_{\pfrak},~~~~~~~{\rm with}~~
    \a_{\pfrak} \equiv \left(\frac{2}{p}\right)(\rmmod~(2+2i)),
  \eeq
  where $\pfrak = (\a_{\pfrak})$.
  Define further the characters $\psi_{6\rmA}$ and $\psi_{6\rmB}$
   on the prime ideals $\pfrak|p$ of $\mathQ(\sqrt{-3})$ dividing
   the rational primes $p$ as
  \bea
   \psi_{6\rmA}(\pfrak) &=& \a_{\pfrak},~~~~~~~~{\rm with}~~
   \a_{\pfrak} \equiv 1(\rmmod~3), \nn \\
   \psi_{6\rmB}(\pfrak) &=& \a_{\pfrak},~~~~~~~~{\rm with}~~
   \a_{\pfrak} \equiv \left(\frac{3}{p}\right)(\rmmod~(2+4\xi_3)).
  \eea
  These characters provide the Hecke theoretic interpretation
  of the Mellin transforms of the Hasse-Weil L-functions of the elliptic curves
  $E^4, E^3$ and $E^6$ respectively. Renaming $E^3$ and $E^6$ as
  $E^{6\rmA}:=E^3$ and $E^{6\rmB}:=E^6$, one can check that
   \beq
    L_{\rmHW}(E^d,s) = L(\psi_d,s),~~~~~~~~d=4,6\rmA, 6\rmB.
   \eeq

 The motivic L-series of $S^4$, $S^{6\rmA}$ and $S^{6\rmB}$
  can then be described as
   \beq
    L_{\Om}(S^d,s) = L(\psi_d^2,s),~~~~~~~~~d=4,6\rmA
   \eeq
   for the first two surfaces, and as
   \beq
    L_{\Om}(S^{6\rmB},s) = L(\psi_{6\rmB}^2\otimes \chi_3,s)
   \eeq
  for the third example. These results provide an alternative demonstration
  that the elliptic curves $E^4$, $E^3$ and $E^6$ can be viewed as the
  modular building blocks of the K3 surfaces considered here. They
  also explain the relations (\ref{3-2relation1}) and
  (\ref{3-2relation2}) by rewriting the coefficients of the Hecke
  L-series of the weight three forms in terms of the real coefficients
  of the weight two forms.

The Hecke interpretation of the $\Om-$motivic L-series of elliptic
curves and K3-surfaces provides a way to explicitly prove the
modularity of these series to all order of their $q-$expansions by
using some results from Hecke and Shimura. Alternatively, it is
possible to point to general modularity result for extremal K3
surfaces defined over number fields \cite{si77,l95}. It is useful,
however, to see modularity explicitly, and a proof is contained in
the appendix.

\subsection{K3 Arithmetic moonshine}

It was pointed out in \cite{s05} that the modular forms given by
the Mellin transform of the Hasse-Weil L-functions of elliptic
Brieskorn-Pham curves can be interpreted as generalized
McKay-Thompson series associated to the largest Mathieu group
$M_{24}$, leading to the notion of what might be called arithmetic
moonshine. It is therefore natural, as an aside, to ask whether
the modular forms that appear in the context of extremal
Brieskorn-Pham K3 surfaces can also be interpreted as such series.
This is the case if the group considered is enlarged from the
Mathieu group to the Conway group, defined as the automorphism
group of the Leech lattice.

The specific group elements associated to the forms encountered in
the present paper are listed in Table 6, where the notation of
ref. \cite{m96} is adopted, which in turn follows the atlas
\cite{atlas}. The numerical part of the group element denotes its
 order while the letter part separates different elements of the
 same order.

 {\small
  \begin{center}
  \begin{tabular}{l| c c c c c c}
 Form   &$\Theta^1_{1,1}(q^3)\Theta^1_{1,1}(q^9)$
        &$\Theta^1_{1,1}(q^6)^2$    &$\Theta^1_{1,1}(q^6)^2\otimes \chi$
        &$\eta^6(q^4)$              &$\eta^3(q^2)\eta^3(q^6)\otimes \chi_3$
        &$\frac{\left(\Theta^2_{1,1}(q)\Theta^2_{1,1}(q^4)\right)^2}{\Theta^2_{1,1}(q^2)}$ \\

\hline
 Conway &$(3D,3B)$
        &$6I$                       &$8F|_T$
        &$4F$                       &$6G|_T$
        &$8E$   \tabroom \\
 class  & & & & &  \\

\hline
\end{tabular}
\end{center}}

\centerline{{\bf Table 6.}~{\it Conway classes for the forms of
Table 5.}}

It is intriguing to see that all geometrically induced conformal
field theoretic modular forms obtained so far admit a sporadic
group theoretic interpretation. It is an open question whether
this relation can help to provide a conceptual foundation of the
relations established so far.

\vskip .1truein

\section{Appendix: modularity proof}

\subsection{Faltings-Serre-Livn\'e strategy}

Modularity of the L-series considered here can be shown in several
ways, e.g. by using their complex multiplication origin and
results of Hecke and Shimura. It is also possible to refer to some
general results of Livn\'e. Calabi-Yau modularity is not
restricted to manifolds with CM, as illustrated by many examples
reviewed in \cite{y03}. For completeness, this appendix contains a
brief modularity proof of the $\Om-$motives of extremal K3
surfaces of Brieskorn-Pham type by using techniques essentially
developed by Faltings and Serre. It was first observed by Faltings
\cite{f83} that different Galois representations can be shown to
be identical if they agree at a finite number of primes. This
observation has been developed by Serre \cite{s85} and Livn\'e
\cite{l87}, and Livn\'e in particular made it into a practical
tool by specifying precisely the set of primes that has to be
tested in order to guarantee agreement of two representations. The
varieties discussed here are defined over $\mathQ$, hence
Livn\'e's more general theorem can be reduced to the following
form, previously considered by Verrill \cite{v00} (see \cite{y03}
for more references in this direction). Let $\mathQ_{\ell}$ be the
$\ell-$adic field and denote by $\rmF_p$ the Frobenius element at
the prime $p$.

{\bf Theorem 5.}~{\it Let $S$ be a finite set of rational primes,
   and denote by $\mathQ_S$ the compositum of all quadratic extensions of
   $\mathQ$ unramified outside of $S$. Suppose }
  \beq
   \rho_1,\rho_2: \rmGal(\omathQ/\mathQ) \lra \rmGL(2,\mathQ_2)
  \eeq
 {\it  are $2-$adic
  continuous representations, unramified outside of $S$, which
  further satisfy the following conditions: \hfill \break
  1) $\rmtr~\rho_1 \equiv \rmtr~\rho_2 \equiv 0 ~\rmmod~2$. \hfill
  \break
  2) There exists a finite set $T\neq \emptyset$ of rational primes,
  disjoint from $S$, for which \hfill \break
   \phantom{wha} a) the image of the set $\{\rmF_p\}_{p\in T}$ in
    $\rmGal(\mathQ_S/\mathQ)$ is surjective, \hfill \break
   \phantom{wha} b) $\rmtr~\rho_1(\rmF_p) = \rmtr~\rho_2(\rmF_p)$
            for all $p\in T$, \hfill \break
   \phantom{wha} c) $\rmdet~\rho_1(\rmF_p) = \rmdet~\rho_2(\rmF_p)$
           for all $p\in T$.
  \hfill \break
  Then $\rho_1$ and $\rho_2$ have isomorphic semi-simplification.}

The idea therefore is to translate the geometric and modular
computations into two Galois representations, and to show
agreement by satisfying the requirements of Livn\'e's result. The
geometric part is given by the representations of the absolute
Galois group $\rmGal(\omathQ/\mathQ)$ on the $\ell-$adic
cohomology group
 \beq
 \rho_1 = \rho_{\ell}^i: \rmGal(\omathQ/\mathQ) ~\lra
  ~\rmAut(\rmH^i_{\acute{e}t}(\bX,\mathQ_{\ell})),
 \eeq
 where $\ell$ is a prime different from the prime of reduction,
 $X/\mathQ$ is assumed to be smooth projective variety,
 $\bX = X\otimes_{\mathQ} \omathQ$, and $\rmH^i_{\acute{e}t}$ indicates
 the $i^{\rmth}$ \'etale cohomology group.

In the absolute Galois group there exists a distinguished element,
the geometric Frobenius endomorphism $\ormF_p \in
\rmGal(\omathQ/\mathQ)$. Given the representation $\rho_{\ell}^i$
these Frobenius elements act on the \'etale cohomology, and it
turns out that the polynomials $\cP^i_p(t)$  that
 define the geometric L-function associated
 to these cohomology groups can be determined in terms of
 these Frobenii as
  \beq
   \cP_p^i(t) =
   \rmdet\left(1-\rho_{\ell}^i(\ormF_p)|_{\rmH^i_{\acute{e}t}(\bX,\mathQ_{\ell})}t
        \right).
   \eeq

The second representation is the 2$-$dimensional Galois
representation $\rho_2 = \rho_f$ associated by Deligne \cite{d71}
to Hecke eigenforms of arbitrary weight, where
 \beq
  \rho_f: ~\rmGal(\omathQ/\mathQ) \lra \rmGL(2,\mathQ_{\ell})
 \eeq
 is a representation that is unramified outside of $\ell$ and the
 prime divisors of the level $N$ of the modular form. This
 $\ell-$adic representation can be realized in the cohomology of certain
 $\ell-$adic sheaves over a modular curve. These can be defined over
 $\mathQ$ and therefore $\rmGal(\omathQ/\mathQ)$ acts on the $\ell-$adic
 cohomology. Deligne has shown that these representations
 have the properties
  \bea
   \rmtr ~\rho_f(\ormF_p) &=& a_p \nn \\
   \rmdet~\rho_f(\ormF_p) &=& p^{w-1}\e(p)
  \eea
  for all primes $p$ different from $\ell$. In the present discussion of
  K3 surfaces the focus is on modular forms of weight three and the goal is to
 use Theorem 5 to prove the identity of the geometric representation and the modular
  representation.

\subsection{Examples}

The application the strategy outlined above to the K3 surface $S^4
\subset \mathP_3$ starts with the determination of the set $T$ of
primes considered in Livn\'e's theorem. This is obtained by
considering a representation of the Galois group of the composite
field $\mathQ_S$. For $S^4$ the bad prime is $p=2$, which is also
the only divisor of the level of  modular form $\eta(4\tau)^6 \in
S_3(\Gamma_0(16), \chi_{-1})$. Hence $S=\{2\}$, and the composite
field is $\mathQ_{\{2\}} = \mathQ(i,\sqrt{2}) = \mathQ(\xi_8)$.
Therefore the set $T$ of primes can be chosen as \beq
  T=\{3,5,7,17\}.
 \eeq
 This set of primes is a subset of the primes for which agreement
 of the modular and the geometric L-series is shown by the
 computations in the previous subsections. It remains to establish
 the conditions formulated in the theorem.

 The fact that $\rmtr~\rho_1 = 0~\rmmod~2$ follows by noting
 that the quartic surface can be constructed via the twist map, as
 explained above.  The elliptic curve $E^4$ involved in the
 construction of $S^4$ admits complex multiplication, and it can
 be shown that its coefficients $a_p$ satisfy $a_p = 0~\rmmod~2$ (see
 e.g. \cite{z00}, or \cite{s05}). Hence the same follows for $\rho_1$.
 Alternatively, this follows from the Jacobi-sum formulation
 of the L-function of the $\Om-$motive.
 The same congruence holds for $\eta^6(4\tau)$, which can be seen
 e.g. via Jacobi's expansion for $\eta^3(\tau)$ \cite{j1829}.
 The determinant condition, finally, follows from Weil's result for Jacobi
 sums in the geometric case, and Deligne's result in the case of
 the modular form.

The case of the degree six surface $S^{6\rmB} \subset
\mathP_{(1,1,2,2)}$ is similar to the discussion of the quartic.
The set of bad primes is $S=\{2,3\}$, which
 gives the set of divisors of the level of the modular form. This
leads to the composite field $\mathQ_{\{2,3\}} =
\mathQ(\xi_{24})$. The set of primes representing the Galois group
of this field is given by
 \beq
  T=\{5,7,11,13,17,19,23,73\}.
 \eeq
 This set of primes is again a subset of the primes considered
 above for the surface $S^6$. Comparison between the geometric and
 modular L-series shows agreement on the set of primes given by
 $T$.

The fact that $\rmtr~\rho_1 =0~\rmmod~2$ can again be seen via the
twist construction. The curve $E^6$, the building block of $S^6$,
has complex multiplication, and the coefficients
 of its Hasse-Weil L-series are zero mod 2. Hence it follows that
 this holds also
 for $\rho_1$. The remaining assumptions of the theorem follow in
 the same way as in the case of the quartic surface in combination
  with results proven in \cite{v00}.

The surface $S^{6\rmA}$ does not satisfy the conditions of
Livn\'e's theorem since the coefficients can be odd. The following
result, proven in \cite{s04}, can be used to complete the proof.

{\bf Proposition.} {\it Let $\rho_1,\rho_2$ be two 2-adic Galois
representations with the same determinant and even trace at
$\rmFr_{11}$ or $\rmFr_{13}$, which are unramified outside
$\{2,3\}$. Then they
 have isomorphic semi-simplifications if and only if for any
 $p \in \{5,7,11,13,17,19,23,31,37\}$ the traces of the Frobenii
 are identical, $\rmtr~\rho_1(\rmFr_p) = \rmtr~\rho_2(\rmFr_p)$.}

A discussion of the notion of semi-simplification can be found in
\cite{t04}. Important here is the implication that the two
L-functions associated to the Galois representations are
identical. It follows from the Jacobi sum representation of the
L-series and Deligne's result that the determinants of the
geometric representation and the modular representations are
identical. The computations collected in Table 1 furthermore show
agreement for the traces at the primes determined by the
proposition. Hence the proof follows.

\vskip .4truein

{\large {\bf Acknowledgement.}} \hfill \break
 It is a pleasure to thank Monika Lynker for discussions, and
 John Stroyls for generously making available his library.
 This work was supported in part by an Incentive Grant for Scholarship
 at KSU, and while the author was a Scholar at the Kavli
Institute for Theoretical Physics in Santa Barbara. It is a
pleasure to thank the KITP for hospitality. This work was
supported in part by the National Science Foundation under Grant
No. PHY99$-$07949.

\end{document}